\documentclass[aps,reprint,superscriptaddress,nofootinbib,showpacs]{revtex4-1}
\usepackage{amsmath,latexsym,amssymb,hyperref,graphicx}
\usepackage{slashed}

\hypersetup{colorlinks,citecolor= nicegreen,linkcolor= nicered}
\usepackage{color}
\definecolor{nicered}{rgb}{0.7,0.1,0.1}
\definecolor{nicegreen}{rgb}{0.1,0.5,0.1}

\newcommand{\beq}{\begin{equation}}
\newcommand{\eeq}{\end{equation}}
\newcommand{\bea}{\begin{eqnarray}}
\newcommand{\eea}{\end{eqnarray}}
\newcommand{\Tr}{{\rm Tr}}

\begin{document}

\title{Self-gravitating non-abelian kinks as brane worlds}

\author{Alejandra Melfo}
\affiliation{Centro de F\'isica Fundamental, Universidad de Los Andes,
M\'erida, Venezuela}
\affiliation{International Centre for Theoretical Physics, 34100 Trieste, Italy}

\author{Roger Naranjo}
\affiliation{Centro de F\'isica Fundamental, Universidad de Los Andes,
M\'erida, Venezuela}

\author{Nelson Pantoja}
\affiliation{Centro de F\'isica Fundamental, Universidad de Los Andes,
M\'erida, Venezuela}

\author{Aureliano Skirzewski}
\affiliation{Centro de F\'isica Fundamental, Universidad de Los Andes,
M\'erida, Venezuela}

\author{Juan Carlos V\'asquez}
\affiliation{Centro de F\'isica Fundamental, Universidad de Los Andes,
M\'erida, Venezuela}

\date{\today}

\begin{abstract}
  \noindent  
We address the properties of self-gravitating domain walls arising from the breaking of  an $SU(N)\times Z_2$- symmetric theory. In  the particular case of $N=5$, we find that the two classes of stable  non-abelian kinks possible in flat space have an analogue in the gravitational case, and construct the analytical solutions.   Localization of fermion fields in different representations of the gauge group in these branes is investigated. It is also shown that non-abelian gauge fields localization cannot be achieved through interactions with the brane, but that in one of the two classes of kinks this localization can be implemented {\it via} the Dvali-Shifman mechanism. 

\end{abstract}


\maketitle

\section{Introduction}

Domain wall solutions of the Einstein-scalar field equations as a regularization of the infinitely thin Randall-Sundrum brane \cite{Randall:1999vf} have been extensively considered in the literature. While preserving the crucial property of localizing gravity \cite{DeWolfe:1999cp,Gremm:1999pj,Kakushadze:2000zn,CastilloFelisola:2004eg}, these thick walls have also some additional  potentially useful properties, arising from the interaction of the scalar field with other fields in the theory \cite{Bajc:1999mh,Kehagias:2000au,Ringeval:2001cq,Koley:2004at,Melfo:2006hh,Liu:2009dw}, or simply from a non trivial internal structure \cite{Melfo:2002wd,Bazeia:2003aw,Guerrero:2005aw,Guerrero:2006gj}. The best known example is the addition of  a Yukawa coupling to the wall's field \cite{Bajc:1999mh} that enables fermions of one chirality to localize on the brane,  compensating for the gravitational effect that repels them. Gauge field localization is more difficult \cite{Huber:2000fh,Dvali:2000rx,Dubovsky:2001pe,Batell:2006dp,Guerrero:2009ac}, as the scalar field is taken invariant under any gauge symmetry, in order to ensure  the stability of the wall through topological arguments. 

However, as pointed out in a series of  papers \cite{Dvali:1997sa,Pogosian:2000xv,Vachaspati:2001pw,Pogosian:2001fm}, it is perfectly possible to build a stable domain wall in flat space out of a scalar field transforming non trivially under a non abelian group. In particular, it is shown in \cite{Vachaspati:2001pw}, that a quartic field theory with a $SU(N) \times Z_2$ symmetry broken down by the vev of an scalar transforming under the adjoint of $SU(N)$ has perturbatively stable domain wall solutions for odd  $N>3$. 

It is only natural to ask if similar solutions can be found for the gravitating case.  Such a thick brane would have interesting properties, as the scalar field will interact directly with the gauge sector, opening up the possibility of localizing the gauge fields on the brane. Inside the wall, a subgroup of the original symmetry would be preserved, allowing for localization of multiplets of fermions that could be of phenomenological interest. 

In order to investigate this possibility, in section \ref{SU_5_branes} we have taken a toy example with the minimal choice $N=5$, and solved the Einstein-scalar field equations for a scalar field transforming in the adjoint, with a sixth-order potential. Two symmetry breaking patterns are considered, $SU(5) \times Z_2 \longrightarrow {SU(3)\times SU(2) \times U(1)}/\left({Z_3\times Z_2}\right)$ and $SU(5) \times Z_2 \longrightarrow {SU(4)\times U(1)}/{Z_4}$, finding in both cases analytical domain wall solutions. We then study fermion localization on these solutions in section \ref{fermion_localization}, by considering its Yukawa interactions with fermion fields in the fundamental representation of $SU(5)$, and showing how the zero modes of some of the fermion components are localized.  

Next, we turned to localization of gauge fields in section \ref{gauge_localization}. First, we address the issue of whether the gauge interactions of the scalar field suffice to do that. We conclude that although a subset of the gauge fields have a localized massless mode on the brane, they necessarily interact with gauge fields that propagate in the bulk unveiling the presence of the extra dimension. This is true for any non-abelian wall arising from the breakdown of $SU(N)\times Z_2$ by a field in the adjoint representation.  We then discuss the possibility of gauge field localization {\it via} the Dvali-Shifman (DS) mechanism \cite{Dvali:1996xe} as a result of the so-called "clash of symmetries" (see e.g. \cite {Davidson:2007cf} and references therein). We find that for the brane with symmetry breaking $SU(5) \times Z_2 \longrightarrow {SU(4)\times U(1)}/{Z_4}$ this mechanism leads to such localization.

 \section{Selfgravitating thick branes in $SU(5) \times Z_2$}\label{SU_5_branes}
 
 Let us consider the $(4+1)$-dimensional theory given by
 \begin{equation}\label{model}
 S=\int d^4 x \,d y \sqrt{-g}\left[ \frac{R}{2} - \text{Tr}(\partial_m \Phi \partial^m \Phi)-V(\Phi)\right]
 \end{equation}
where $R$ is the scalar curvature, $g$ is the determinant of the metric, $\Phi$ is a field that transform in the adjoint representation of $SU(5)$ and $V(\Phi)$ is a potential invariant under the action of this group.\footnote{We use units were $\hbar = G = c = 1$} 

Now, since  the elements of the center of $SU(N)$  transform $\Phi$  by  multiplicative factors of $\exp({{2\pi i}/{N}})$, for even $N$ the transformation $\Phi \rightarrow -\Phi$ is included in the center, while for odd values of $N$ it is not \cite{Vachaspati:2001pw}. It follows that for $N=5$, in order to generate topological kink solutions that break the discrete $Z_2$ symmetry, the symmetry of (\ref{model}) must be chosen as $SU(5) \times Z_2$. 

It has been shown in \cite{Vachaspati:2001pw} that 
  in flat space and for a $SU(5)$-invariant fourth-order potential,  the  vev of $\Phi$ will form perturbatively stable domain walls when the symmetry breaking pattern is 
 \begin{equation}
 SU(5) \times Z_2 \longrightarrow \frac{SU(3)\times SU(2) \times U(1)}{Z_3\times Z_2}
 \end{equation}
provided that the vevs of $\Phi$ at spatial infinity are related  both by a $Z_2$ transformation and an $SU(5)$ one, i.e 
\beq
\Phi(y=-\infty)=-U\Phi(y= +\infty)U^{-1}  
  \label{boundarycondition}
\eeq
where  $U$ is an element of  $SU(5)$ and $\Phi$ depends only on the coordinate $y$. 

On the other hand,  for a fourth-order potential with a $Z_2$ symmetry, it is well known that there is only one other possible minimum, giving rise to the    symmetry-breaking pattern
 \begin{equation}
 SU(5) \times Z_2 \longrightarrow \frac{SU(4) \times U(1)}{Z_4}
 \end{equation}
Along the lines of \cite{Vachaspati:2001pw}, one can find a fourth-order potential for $\Phi$ such that 
 imposing boundary conditions as in \eqref{boundarycondition} domain walls can be formed in this case also.   It is straightforward to prove that these walls are stable under small perturbations.    
 
Each of these two patterns of symmetry breaking provides only one class of stable kinks, other solutions being either unstable or equivalent. 
  
Having thus identified the only two possibilities of constructing a stable kink from an $SU(5)\times Z_2$ theory, we turn now to the issue of incorporating gravity. In order to obtain analytical kink solutions, the potential must be chosen as a sixth-order one. Kink solutions for a sixth-order potential for a single scalar field coupled to gravity have been found in \cite{DeWolfe:1999cp}.  For the $SU(5)\times Z_2$ case we begin by writing the most general potential
\begin{align}
V(\Phi) &=-m^2 \text{Tr}[\Phi^2]+h (\text{Tr}[\Phi^2])^2+\lambda \text{Tr}[\Phi^4] + \alpha (\text{Tr}[\Phi^2])^3 
\nonumber \\
&+ \beta (\text{Tr} [\Phi^3])^2+ \gamma (\text{Tr} \Phi^4)(\text{Tr} \Phi^2)+V_0   \label{mipotencial}
\end{align}
where $V_0$ is a constant to be fixed.

Next, we follow a similar strategy as in  \cite{Vachaspati:2001pw}, factorizing
\begin{equation}
\Phi=\phi_M(y) \mathbf{M} + \phi_P (y) \mathbf{P} \label{misolucion}
\end{equation}
where \text{M} and \text{P} are two orthogonal diagonal generators of $SU(5)$. We demand that they satisfy the   conditions 
\bea
\Tr(\mathbf{M}^2)=\Tr(\mathbf{P}^2) &=& 1/2 \nonumber \\
\Tr(\mathbf{M}\mathbf{P})=\Tr( \mathbf{M}^3 \mathbf{P}) = \Tr(\mathbf{P}^3 \mathbf{M}) &=& 0  \nonumber \\
\Tr (\mathbf{P}^3) \Tr (\mathbf{M} \mathbf{P}^2)=  \Tr (\mathbf{M}^3) \Tr (\mathbf{M}^2 \mathbf{P})  &=& 0 \nonumber  \\
\Tr(\mathbf{M} \mathbf{P}^2) \Tr(\mathbf{P} \mathbf{M}^2) =\Tr (\mathbf{P}^3) \Tr (\mathbf{M}^3) &=& 0
\eea

Analytic solutions to the Einstein-scalar field system can be found, as in the case without gravity, if the cross-terms in (\ref{mipotencial}) containing  both $\phi_M$ and $\phi_P$ vanish. This requires the couplings to be related by 
\bea \label{conditions}
 h &=& - 12 \Tr(\mathbf{P}^2 \mathbf{M}^2) \,  \lambda \nonumber \\
 \alpha &=& \frac{4}{3}\left[ \frac{ \Delta_2 \Tr (\mathbf{P}^4) -\Delta_1 \Tr (\mathbf{M}^4) }{ \Delta_1 - \Delta_2}   -  6 \Tr(\mathbf{P}^2 \mathbf{M}^2) \right] \, \gamma \nonumber \\
  \beta &=&  \frac{1}{6}\left[  \frac{\Tr (\mathbf{M}^4) - \Tr (\mathbf{P}^4)}{ \Delta_1 - \Delta_2}  \right] \, \gamma
\eea
where
\bea
  \Delta_1& \equiv & 3 [\Tr (\mathbf{P}^2 \mathbf{M})]^2 + 2 \Tr (\mathbf{P}^3) \Tr (\mathbf{M}^2 \mathbf{P}) \nonumber \\
    \Delta_2 & \equiv&  3 [\Tr (\mathbf{P} \mathbf{M}^2)]^2 + 2 \Tr (\mathbf{M}^3) \Tr (\mathbf{P}^2 \mathbf{M})
\eea 
and $\Delta_1, \Delta_2$ are required  not to be zero simultaneously. 
With  $\Phi$ of the form given in \eqref{misolucion} and the space-time metric {\it ansatz}
\begin{equation}
ds^2=e^{2A(y)}\eta_{\mu\nu}dx^{\mu} dx^{\nu}+dy^2;  \quad \mu,\nu=0,1,2,3; \label{mimetrica}
\end{equation}
where $\eta_{\mu\nu}$ is the four-dimensional Minkowski metric,
the Einstein-scalar field equations for this system are
\begin{align}
&3A''=-(\phi_M^{'2}+\phi_P^{'2}) \nonumber  \\
&\frac{3}{2}A''+6A'^2=-V(\Phi_k)  \nonumber \\
&\phi_M^{''}+4A'\phi_M^{'}= -\frac{\partial V}{\partial \phi_M }   \nonumber \\
&\phi_P^{''}+4A'\phi_P^{'}=-\frac{\partial V}{\partial \phi_P }  
\label{einseq}
\end{align}
where the primes indicate differentiation with respect the additional coordinate.

Now we impose trivial and non-trivial boundary conditions for $\phi_P$ and   $\phi_M$  respectively
\begin{align}
&\phi_P(+\infty)=\phi_P(-\infty) \label{C2} \\
&\phi_M(+\infty)=-\phi_M(-\infty) \label{C1} 
\end{align}   
 
It is now possible to find solutions to  \eqref{einseq}, provided there is still one more relation between the parameters in the potential and  the constant $V_0$ is appropriately chosen. 
Let us write the potential as
\bea
V & =& -\frac{m^2}{2} \phi_M^2 + \frac{ \lambda_M}{4} \, \phi_M^4 +  \frac{\gamma_M}{6} \, \phi_M^6 \nonumber \\
 & & -\frac{m^2}{2} \phi_P^2 + \frac{ \lambda_P}{4} \, \phi_P^4 +  \frac{\gamma_P}{6} \, \phi_P^6  + V_0
\eea
where the couplings $\lambda_M, \lambda_P, \gamma_M$ and $\gamma_P$ are calculated after imposing  the decoupling conditions (\ref{conditions}). Then
solutions are found to be 
\begin{equation}
\Phi(y)= v \text{ }[\text{tanh}(by) \mathbf{M} +\kappa \mathbf{P}] 
\label{pared}
\end{equation}
with a warp factor
\beq
 A(y)=-\frac{v^2}{9}[2 \ln \left(\text{cosh}(by) \right)+\frac{1}{2}\tanh^2(by)]
 \label{warp}
\eeq
where 
\beq
v^2 = \frac{1}{4}\left( \frac{\lambda_M}{\gamma_M} - \frac{9}{2}\right) \, ; \; b^2= \frac{9}{4}\, v^2 \,\gamma_M
\eeq
provided that
\bea
m^2 &=& 2 b^2\left( 1 + \frac{2}{3} v^2\right) \\
V_0 &=& -  v^4 b^2 \frac{8 }{27}  + \kappa^4 v^4\left(\frac{\lambda_P}{4} + \frac{\gamma_P}{3} \kappa^2 v^2 \right) 
\label{masscond}
\eea

Here  a crucial role is played by the constant   $\kappa$,  given by
\beq
\kappa^2= \frac{\lambda_P}{\lambda_M}\frac{\gamma_M}{\gamma_P}\frac{ 1 + \sqrt{1 + 4 m^2 \gamma_P/\lambda_P^2}   }{1 + \sqrt{1 + 4 m^2 \gamma_M/\lambda_M^2} }
\eeq
In order for the solutions to exist, $\kappa$ must be a numerical factor, and this depends on the choice of  $\mathbf{M}$ and $\mathbf{P}$. 

The potential $V(\Phi)$ takes the asymptotic values
\begin{equation}
V(\Phi)_{y=-\infty}=V(\Phi)_{y=+\infty}=-\frac{8}{27}b^2  \, v^4
\end{equation}    
so that the space-time is asymptotically $AdS_5$.

\subsection{A thick brane with symmetry breaking $SU(5) \times Z_2 \longrightarrow {SU(3)\times SU(2) \times U(1)}/\left({Z_3\times Z_2}\right)$}\label{SU(5)toSU(3)XSU(2)XU(1)}

In order to satisfy the boundary conditions  \eqref{boundarycondition} and ensure stability, the first class of kinks has $\Phi$ taking the asymptotic values
\bea \label{boundaryI}
\Phi_{\bf A}(+\infty)& \sim &  v\,{\rm diag( \;3, \;3,-2,-2,-2)} \nonumber \\
\Phi_{\bf A}(-\infty) & \sim &  v\,{\rm diag( \;2, \;2,-3,-3,\, \;\;2)}
\eea
 
We  take  {\bf M} and \text{ \bf P} as
\bea
\mathbf{M}_{\bf A} &=&\frac{1}{\sqrt{40}} {\rm diag (1,1,1,1,-4)}
\nonumber \\
\mathbf{P}_{\bf A} &=&\frac{1}{2\sqrt{2}}  {\rm diag (1,1,-1,-1,0)}
\eea
 
Then solutions of the form  (\ref{pared} -\ref{masscond}) exist  and have
\beq
\kappa_{\bf A} = \sqrt{5}
\eeq 
so that  they satisfy the boundary conditions (\ref{boundaryI}). This solution is the gravitational analogue of the wall found in \cite{Vachaspati:2001pw}.

Let us briefly review the symmetry breaking induced by this wall. At spatial infinity along the bulk coordinate, the remaining gauge symmetries $H$ are the same, but correspond to different embeddings in $SU(5)$ as $y\rightarrow -\infty$ or $y\rightarrow\infty$. We have at $y\rightarrow\pm \infty$
\beq
H^{\bf A}_\pm = \frac{SU(3)_\pm \times SU(2)_\pm \times U(1)_\pm}{Z_2\times Z_3}
\eeq
Inside the wall where the vev in the {\bf M} direction vanishes, the unbroken group $H_0$ is
 \beq
 H^{\bf A}_0 = \frac{SU(2)_+ \times SU(2)_- \times U(1)_M \times U(1)_P}{Z_2 \times Z_2}
 \eeq
where we have named the $U(1)$ factors according to their generators. The embedding is such that $SU(2)_\pm \subset SU(3)_\mp $. 

\subsection{A thick brane with symmetry breaking $SU(5) \times Z_2 \longrightarrow {SU(4)\times U(1)}/{Z_4}$}\label{SU(5)toSU(4)}

 A brane with $\Phi$ taking the asymptotic values
 \bea \label{boundaryII}
\Phi_{\bf B}(+\infty)& \sim & v\,{\rm diag( \;\;1, \; \;1, \;\;1, \;\;1,-4)} \nonumber \\
\Phi_{\bf B}(-\infty) & \sim & v\,{\rm diag( -1, -1,-1,4, -1)}
\eea
can be found by choosing 
\bea
\mathbf{M}_{\bf B} &=&\frac{1}{\sqrt{60}} {\rm diag (-2,-2,-2,3,3)}
\nonumber \\
\mathbf{P}_{\bf B} &=&\frac{1}{2}  {\rm diag (0,0,0,1,-1)}
\eea
where now
\beq
\kappa_{\bf B} = \sqrt{\frac{5}{3}}
\eeq 

An analogue of this wall for the flat space case (with a fourth-order  potential) exists, although it has not been reported to our knowledge. It is not difficult however to show that it is perturbatively stable in the flat space case. It is also straightforward to show  that walls built by choosing boundary conditions with different embeddings of the $SU(4)\times U(1)$  group (i.e., interchanging the position of the elements on the diagonal of, say, $\Phi_{\bf B}(-\infty)$)  are equivalent to this one, corresponding to the same class of kinks. 

In this case, we have at spatial infinity
 \beq
 H^{\bf B}_\pm = \frac{SU(4)_\pm \times  U(1)_\pm}{Z_4}
\eeq
and inside the wall
 \beq
 H^{\bf B}_0 = \frac{SU(3) \times U(1)_M \times U(1)_P}{Z_3 }
 \eeq
The $SU(3)$ group in $H^{\bf B}_0$ is embedded in distinct manners in the $SU(4)$ groups of $H^{\bf B}_+ $ and $H^{\bf B}_- $ .

\section{Fermion localization}\label{fermion_localization}

As is well known \cite{Bajc:1999mh}, fermions in a  warped geometry are not localized on the brane, unless a Yukawa coupling with the scalar field that generates the brane forces them. In order to see what happens in the non-abelian case, we  take as an example fermions in the fundamental representation of $SU(5)$. 

Consider the  5-dimensional action in a warped spacetime with metric (\ref{mimetrica}), for a massless spinor field $\Psi$ in the fundamental representation of $SU(5)$ coupled  to a scalar field $\Phi $ in the adjoint  
\begin{equation}
\mathcal{S}_{\Psi}= \int d^4x dy \sqrt{-g} \bar{\Psi}^{I}\left[i\Gamma^{m}\delta^{IJ}\nabla_{m}+\eta\Phi^{IJ}\right]{\Psi^{J}}
\end{equation}
where $I,J$ are $SU(5)$ indices, $m,n,\ldots$ are five-dimensional Lorentz indices,  $g$ is the determinant of the metric, $\nabla_{m}$ is the covariant derivative and the $\Gamma^m$  satisfy the Clifford algebra. Dirac's equation reads 
\begin{equation}
\left [\left( i\Gamma^{\mu}e^{-A}\partial_{\mu}+i\Gamma^{y}\left ( 2A+\partial_{y} \right) \right)\delta^{IJ}-\eta\Phi^{IJ}\right]\Psi^J=0
\end{equation}

We now factorize  the fermion field as usual
\beq
\Psi^I(x,y)=  [\psi(x)f(y) ]^I
\eeq
and require that the four dimensional part $\psi^I$ satisfies the Dirac equation for a massless spinor, $\gamma^{\mu}\partial_{\mu}\psi^I=0$. For $\Phi$  given by  \eqref{pared}, the solution for the chiral modes is straightforward 
\begin{equation}
f_{\mathcal{L},\mathcal{R}}^I \propto \exp \left[ -2A(y) \mp \eta \, v \left( Q^I_M \,  \ln(\cosh b y)/b +  \kappa\, Q^I_P\,  y \right)  \right] \label{solucionfermion}
\end{equation}
where  $Q^I_M, Q^I_P$ are the charges of the $I$ component of the fermion multiplet  under $M$ and $P$ respectively. 

 Notice that the term proportional to $\kappa$ can spoil localization as it diverges in either positive or negative infinity. An asymptotic analysis  reveals that localization takes place only for components satisfying
\beq
\left| Q^I_M  \right| > \kappa \,  \left| Q^I_P  \right|
\eeq
Once this equation is satisfied, the sign of $Q_M(\Psi^a)$ will determine which chiral mode is localized for a given choice of the sign of the Yukawa coupling, provided it is large enough.

Take as an example solution {\bf A}. Under the brane's symmetry group $H^{\bf A}_0$, the fermion field decomposes into a $SU(2)_+$  doublet $f^+$, a $SU(2)_-$ doublet $f^-$, and a singlet $f^0$ under both.   Only (one chiral mode of)  the singlet  component $f^0$ is localized on the brane, provided that
\begin{equation}
| \eta |  > \frac{\sqrt{10}}{9} \, v \, b
\end{equation}

As another example, consider a fermion field transforming under the  the ${\bf 10}$-dimensional representation of $SU(5)$: it will have one chiral mode of four of its components confined in case {\bf A}, corresponding to the bidoublet under the two SU(2) factors, which is not charged under $\mathbf{P_{\bf A}}$.   In case {\bf B},   the  $SU(3)$ triplet  in the fundamental gets confined.

\section{Gauge field localization}\label{gauge_localization}

Now, let us consider the gauge field sector in the background of a selfgravitating non-Abelian domain wall generated by $\Phi$
 \begin{equation}\label{gauge_action}
S_A=\int dx^4dy \sqrt{-g} \left(  -\frac{1}{4}\text{Tr}\left(F_{m n}\right)^2  - \texttt{g}^2 \text{Tr}([A_m,\Phi])^2  \right)
\end{equation}
where $F_{mn}=\partial_mA_n-\partial_nA_m + [A_m,A_n]$, $A_m$ are the gauge fields and $\texttt{g}$ is the gauge coupling constant. We want to assess whether the interactions with the scalar field $\Phi$ can help to localize on the brane the massless zero modes of 5-dimensional gauge fields $A_m$. Since we are interested in the gauge field propagation in the additional dimension, we need to consider only the quadratic terms in the action. The effect of the terms of third and fourth order in the field $A_m$ will be discussed later. 

The linear part of the equation of motion for the gauge fields $A_m$ is
\begin{equation}
\frac{1}{\sqrt{-g}} \partial_m \left( \sqrt{-g}g^{mp}g^{nq}(\partial_qA^a_p-\partial_pA^a_q)\right) - \left(M^2\right)^{ab}g^{np}A_{p}^b=0 \label{ecuacioncalibre}
\end{equation}
where the mass matrix $M^2$ is given by
\begin{equation}\label{y_mass}
\left(M^2\right)^{ab}=-Tr\left([T^a,\Phi][T^b,\Phi]\right)
\end{equation}
Notice that the symmetry breaking induced by $\Phi$ resembles the proposal of Ref.\cite{Huber:2000fh} (see also \cite{Batell:2006dp}) to localize zero mode gauge fields.

From \eqref{ecuacioncalibre}, in the axial gauge $A_y=0$, we find
\begin{eqnarray}
-e^{-2A} \Box A_{\mu}^a+ e^{-2A}\eta^{\alpha \beta}\partial_{\beta}\partial_{\mu}A_{\alpha}^a 
 - 2A' \partial_{y}A_{\mu}^a &\nonumber\\- \partial_{y}\partial_{y}A_{\mu}^a-\left(M^2\right)^{ab}A_{\mu}^b &=0   \end{eqnarray} and
\begin{equation}
\eta^{\mu\nu}\partial_{\nu}\partial_yA_{\mu}^a=0
\end{equation}
We now factorize  $A_{\mu}^a=\left(A_{\mu}(x)f(y)\right)^a$,  and require that the four-dimensional part $A_{\mu}^a$ satisfies a massless Proca equation, i.e. $\Box A_{\mu}^a=0$ and $\eta^{\mu\nu}\partial_{\nu}A_{\mu}^a=0$ \cite{Huber:2000fh}. Hence
 \begin{equation}\label{componenty}
\left[\left(- \partial_y\partial_y-2A'\partial_{y}\right)\delta^{ab}-\left(M^2\right)^{ab}\right]f(y)^b =0
\end{equation}
From (\ref{y_mass}), it follows that the mass terms depends on the direction 
in $SU(5)$ space.

Let $(M^2)^{(n)}$ be the restriction of $M^2$ along a subset of generators. Along the $n_{\rm br}$ generators  $\{T_{\rm br}\}$ that satisfy
\beq
[T_{\rm br}, \mathbf{M} ] = 0 ,\qquad  [T_{\rm br}, \mathbf{P} ] \neq 0 
\eeq 
the mass term is   constant 
\begin{equation}
(M^2)^{({\rm br})}= c_1\texttt{g}^2\, v^2  \,\textsf{1}_{n_{\rm br}\times  n_{\rm br}}   \label{masaconstante3}
\end{equation}
These generators are broken everywhere, and we have $n_{\rm br} = 8$ for solution {\bf A}, $n_{\rm br}=2$ for solution {\bf B}.

Along the $n_K$ generators $\{K_\pm\}$ defined as those satisfying
\beq
[K_\pm , \mathbf{M} ] =  \mp \, \kappa \, [K_\pm, \mathbf{P} ] \neq 0 
\eeq
it is  given by
\begin{equation}
(M^2)^{(\pm)}= c_2 \texttt{g}^2 \, v^2 \left(   \tanh(by) \pm 1 \right)^2  \,\textsf{1}_{n_K\times  n_K}   \label{masanoconstante1}
\end{equation}
with $n_K = 8 $ for solution {\bf A} and $n_K = 12 $ for solution {\bf B}.
Here $c_1,c_2$ are numerical factors.  For all other generators in the solutions found, $M^2$ is zero.

Localization of gauge fields on this brane can be readily established employing the asymptotic values for the warp factor $A$ and the mass matrix $M^2$ as $y\rightarrow \pm\infty$. For a constant mass, as given by \eqref{masaconstante3}, we find that a normalizable solution exists which asimptotically behaves as
\begin{equation}\label{calibre1}
f(y)^{\rm br}\sim \frac{1}{\sqrt{\alpha}} \,  e^{-\alpha |y|},\qquad y\rightarrow \pm\infty 
\end{equation}
where
\begin{equation}
\alpha = \frac{2}{9}v^2\, b \left( 1 +\sqrt{ 1 - \frac{81 \, c_1 }{4\, b}\texttt{g}^2}  \right)
\end{equation}
On the other hand, for mass terms as given by \eqref{masanoconstante1} there are no normalizable solutions.

 Thus, only vector fields with a constant mass in five dimensions, i.e. vector fields along the  broken generators $T_{\rm br}$, can have a normalizable zero mode on the brane. But the set $\{T_{\rm br}\}$ is not an algebra. Hence, we have that all these normalizable zero modes, although localized on the brane, interact with fields propagating in the bulk {\it  via} third and fourth order terms in (\ref{gauge_action}).
 
There is however another possibility. The  solutions {\bf A} and {\bf B} give non-trivial examples of the so-called  ``clash of symmetries' idea \cite {Davidson:2007cf} to implement the DS mechanism \cite{Dvali:1996xe} for gauge field localization. There are two different embeddings of non-abelian, confining groups outside the brane, so that the abelian gauge fields inside the brane would need to form  massive glueballs in order to escape into the bulk. Non-abelian fields inside the brane would in principle have to form glueballs of a different group, so they can also become  localized. 

However, it is the specific nature of the breaking of $SU(5)$ that determines the possibility of obtaining localization via this mechanism. In  case {\bf A}, the  solution that breaks $SU(5) \times Z_2$ to ${SU(3)\times SU(2) \times U(1)}/\left({Z_3\times Z_2}\right)$ asymptotically, the  $SU(2)_\pm$ factors of $H_0$  are embedded in the $SU(3)_\mp$ part of $H_\mp$. Hence, $SU(2)$ glueballs inside the wall can freely escape to positive or negative infinity. The  $U(1)_M$ and $U(1)_P$ gauge fields  also have components in the direction of the $U(1)\pm$ generators. This renders the DS mechanism useless here. 
 
 Case  {\bf B}  in turn has an $SU(3)$ group inside the wall, embedded (in a different manner) in the two $SU(4)$ groups outside it. It is an ideal framework for the DS mechanism of gauge localization. In this respect it is similar to the example given in   \cite{Davidson:2007cf}, where a non-abelian brane in flat space breaks $E_6$ to $SU(5)\times U(1) \times U(1) $ within the wall. However the (perturbative) stability in this  $E_6$ brane  is rather difficult to ascertain, and its extension to the gravitational case an open question.  
 The breaking $SU(5) \times Z_2 \longrightarrow {SU(4)\times U(1)}/{Z_4}$ gives a simpler example of a stable solution allowing for implementation of the DS idea, and as we have shown, its extension to the case including gravity is straightforward.

\section{Summary and Outlook}

We have shown that it is possible to find analytical solutions to the Einstein-scalar field equations representing a domain wall in five dimensions generated by the vev of a scalar field transforming non trivially under a non-abelian symmetry. We have found such solutions in the toy model case of the scalar being in the adjoint of $SU(5)$, with a sixth-order potential that has also a $Z_2$ symmetry, for the symmetry breakings $SU(5) \times Z_2 \longrightarrow {SU(3)\times SU(2) \times U(1)}/\left({Z_3\times Z_2}\right)$ and $SU(5) \times Z_2 \longrightarrow {SU(4)\times U(1)}/{Z_4}$.  

These walls are capable of confining the zero mode of some components of  fermions transforming as $SU(5)$ multiplets,   through a Yukawa term.  We have determined how localization depends on the charge of these components under  the generators that define the wall. 

Non-abelian gauge fields do not seem to localize on this brane  through interactions with the symmetry breaking non-abelian domain wall. We have shown that although a normalizable zero mode exists for some of these vector fields, these correspond to the generators of the broken symmetries in all space. Furthermore, they necessarily interact with the remaining vector fields that propagate in the bulk, unveiling the existence of the extra dimension. It remains to be seen the effect of brane localized kinetic terms for the gauge fields, as in  \cite{Dvali:2000rx} (see also \cite{Guerrero:2009ac}). 

Finally, we have shown that the nature of the breaking of $SU(5)$ determines the possibility of obtaining gauge field localization via the DS mechanism. We find that in the brane with symmetry breaking $SU(5) \times Z_2 \longrightarrow {SU(4)\times U(1)}/{Z_4}$ this mechanism leads to such localization, while in the one which breaks $SU(5) \times Z_2$ to ${SU(3)\times SU(2) \times U(1)}/\left({Z_3\times Z_2}\right)$ asymptotically, this mechanism cannot be realized.

Several important issues need further investigation. The first is naturally the stability of these solutions. Since to our knowledge, the definition of a topological charge in the non-abelian case is an unsolved question, their perturbative stability (in which the form of the potential plays a key role) should be proved. Stability under small perturbations of various $SU(5)$ kinks in flat space has been analyzed in \cite{Vachaspati:2001pw,Pogosian:2001fm}, we will  address the much more involved case including gravity in a forthcoming publication.

One can also worry as to what extent these toy models can be useful in a more realistic case. Namely, if one wishes to construct a non-abelian braneworld, the symmetry group of the gauge fields localized on the brane should be more akin to the group of the standard model.  This issue is also under investigation.

\section{Acknowledgements} We wish to acknowledge discussions with Francisco J. Rodr\'iguez, who joined us in the beginnings of this research. 
This paper is dedicated to his memory.

\end{document}